\documentclass[fleqn,twoside]{article}
\usepackage{espcrc2}

\usepackage{graphicx}
\usepackage[figuresright]{rotating}

\newcommand{\AmS}{{\protect\the\textfont2
  A\kern-.1667em\lower.5ex\hbox{M}\kern-.125emS}}

\title{Neutralino-stop coannihilation in the CMSSM}

\author{Y. Santoso\address[TPI]{Theoretical Physics Institute, University of
        Minnesota, \\ 
        Minneapolis, MN 55455, USA}%
        \thanks{I would like to thank J. Ellis and K.A.
	Olive, with whom this work was done in collaboration. This work was
	supported in part by DOE grant
	  DE-FG02-94ER-40823.}
        }
       
\begin{document}
\begin{titlepage}
\pagestyle{empty}
\baselineskip=21pt
\vskip -0.5in
\rightline{hep-ph/0205026}
\rightline{UMN--TH--2051/02}
\rightline{TPI--MINN--02/11}
\vskip 0.2in
\begin{center}
{\large{\bf Neutralino-Stop Coannihilation in the CMSSM}~\footnote{Talk
presented at Dark Matter 2002, February 20 - 22, 2002, Marina del Rey, CA, to be
published in the proceeding.}}
\end{center}
\begin{center}
\vskip 0.2in
{{\bf Yudi Santoso}}\\
\vskip 0.1in
{\it
{Theoretical Physics Institute,
University of Minnesota, Minneapolis, MN 55455, USA}}\\
\vskip 0.2in
{\bf Abstract}
\end{center}
\baselineskip=18pt \noindent
Neutralino as the lightest supersymmetric particle (LSP) is a
candidate for supersymmetric dark matter. It is known that coannihilation
effects
could be important in neutralino relic density calculation.
Here we present some results on neutralino-stop coannihilation in the CMSSM. In
this model, the stop $\tilde{t}_1$ can be degenerate with the lightest
neutralino
$\chi$ when $|A_0| \neq 0$ and large, more specifically when  the lighter stop
mass is suppressed by large off-diagonal terms in the stop square mass
matrix. In the region where  $\tilde{t}_1$  is slightly heavier than $\chi$,
the coannihilation effect brings the relic density into the range favored by
cosmology.  While the $\chi
- \tilde{t}_1$ coannihilation channels do not increase the range of $m_{\chi}$,
they do extend the cosmologically preferred region to larger values of $m_0$.

\vfill
\end{titlepage}

\begin{abstract}
Neutralino as the lightest supersymmetric particle (LSP) is a
candidate for supersymmetric dark matter. It is known that coannihilation
effects
could be important in neutralino relic density calculation.
Here we present some results on neutralino-stop coannihilation in the CMSSM. In
this model, the stop $\tilde{t}_1$ can be degenerate with the lightest
neutralino
$\chi$ when $|A_0| \neq 0$ and large, more specifically when  the lighter stop
mass is suppressed by  large off-diagonal terms in the stop square mass
matrix. In the region where  $\tilde{t}_1$  is slightly heavier than $\chi$,
the coannihilation effect brings the relic density into the range favored by
cosmology.  While the $\chi
- \tilde{t}_1$ coannihilation channels do not increase the range of $m_{\chi}$,
they do extend the cosmologically preferred region to larger values of $m_0$.

\vspace{1pc}
\end{abstract}

\maketitle

\section{INTRODUCTION}
A supersymmetric model with conserved $R$-parity provides a candidate for dark
matter particle, i.e. the lightest supersymmetric particle (LSP) is stable and
can sustain the time from the big bang to the current time. In the minimal
supersymmetric standard model (MSSM), all fermion and gauge
fields of the Standard Model are given supersymmetric partners and there are
two Higgs multiplets. Furthermore, supersymmetry (SUSY) is broken and the
breaking is represented by soft breaking terms.
Assuming the most general MSSM, however, one has too many free parameters to
deal with. The number of free parameters is reduced significantly when one
assumes some kind of SUSY breaking mechanism. One such model is the constrained
MSSM (CMSSM), or also known as the minimal supergravity model
(mSUGRA)~\cite{msugra}, 
where some universalities are imposed in the soft breaking terms, such that  
 we have only five free parameters in this model: $m_0$
(the universal scalar mass at the GUT scale), $m_{1/2}$ (the universal gaugino
mass at the GUT scale), $A_0$ (the universal trilinear couplings at the GUT
scale),
$\tan \beta$ (the ratio of the two Higgs vacuum expectation values), and ${\rm
sign}(\mu)$ where $\mu$ is the Higgs mixing parameter.   

The best candidate for dark matter in the CMSSM is the lightest neutralino
$\chi$. In our analysis, we take its relic density to be
$
0.1 \leq \Omega_\chi h^2 \leq 0.3
$, 
where $\Omega = \rho/\rho_{\rm critical}$ and $h$ is the Hubble constant in the
unit of 100~km~Mpc$^{-1}$~s$^{-1}$.
The upper limit is imposed by the measurement of the age of the universe, while
the lower limit is suggested by astrophysical observations. One can relax
the lower limit if there are other sources of dark matter. 

The cosmologically favored regions in the CMSSM have been studied intensively
for more than a decade. This includes the bulk~\cite{bulk1,bulk2}, the $H$ and
$A$ pole rapid 
annihilation~\cite{bulk1}, the neutralino-stau ($\chi - \tilde{\tau}$)
coannihilation~\cite{stau} and the focus point~\cite{focus}
regions. It has also been realized that stop $\tilde{t}_1$ can be degenerate
with $\chi$ when $|A_0|$ is large~\cite{stop0}. Therefore we should also
consider 
the neutralino-stop coannihilation~\cite{stop,eos} to complete our study on the
CMSSM parameter space~\cite{eos2,others}.

\section{NEUTRALINO-STOP COANNIHILATION}

\begin{table*}[htb]
\caption{The neutralino-stop coannihilation channels.}
\label{table:1}
\newcommand{\m}{\hphantom{$-$}}
\newcommand{\cc}[1]{\multicolumn{1}{c}{#1}}
\renewcommand{\tabcolsep}{2pc} 
\renewcommand{\arraystretch}{1.2} 
\begin{tabular}{@{}ll}
\hline
Initial State &  Final States \\
\hline
${\tilde t}_1 {\tilde t}^*_1$ & $g g$, $\gamma g$, $Z g$, $t
{\bar t}$, $b {\bar b}$, $q {\bar q}$, $g h$, $g H$, $Z h$,  
 $Z H$, $Z A$, $W^\pm H^\mp$, $h h$, $h H$, $H H$, $A A$, \\  
~ & $h A$, $H A$, $H^+ H^-$, $l {\bar l}$, $W^+ W^-$, $Z
\gamma$, $ZZ$ \\
${\tilde t}_1 {\tilde t}_1$ & $t t$ \\
$\chi {\tilde t}_1$ & $t g$, $t Z$, $b W^+$, $t H$, $t h$, $t A$, $b H^+$, 
$t \gamma$ \\
${\tilde t}_1 {\tilde \ell}$ & $t \ell$, $b \nu $ \\
${\tilde t}_1 {\tilde \ell}^*$ & $t {\bar \ell}$ \\
\hline
\end{tabular}\\[2pt]
\end{table*}

\begin{figure}[htb]
\vspace{9pt}
\includegraphics[height=15pc]{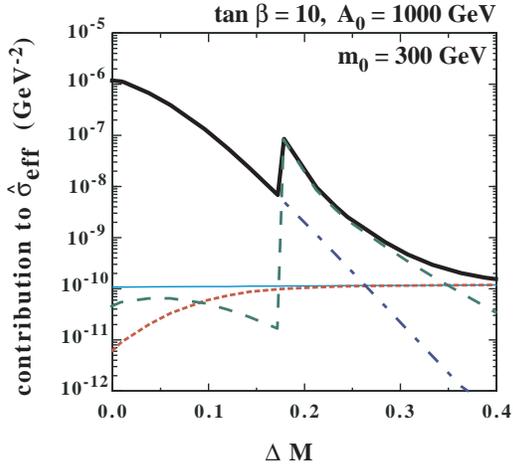}
\caption{The separate contributions to the total effective cross sections
$\hat\sigma_{\rm eff}$ for $x=T/m_\chi=1/23$, as functions
of $\Delta M\equiv(m_{\tilde{t}}- m_\chi)/m_\chi$, obtained by varying $m_{1/2}$, 
with  $A_0 = 1000$~GeV, $\tan \beta = 10$, $\mu > 0$ and 
$m_0 = 300$~GeV. The dark (black) solid line is the total $\sigma_{\rm eff}$.
The (blue) dot-dashed line is the contribution from $\tilde{t}_1 - \tilde{t}_1$
and $\tilde{t}_1 - \tilde{t}_1^{\ast}$. The (green) dashed line is the
contribution from $\tilde{t}_1 - \chi$. The (red) dotted line is the
contribution from $\chi - \chi$. The pale (blue) solid line is the $\chi -\chi$
annihilation cross section if we neglect the coannihilation effect.}
\label{fig:contribution}
\end{figure}

Neutralino relic density is calculated~\cite{relic} by assuming that in the
early universe
neutralino was in thermal equilibrium with other particles. When the expansion
rate of the universe was greater than the scattering rate that keeps the
neutralino in equilibrium, the neutralino froze out. The relic density is
obtained by
evolving the Boltzmann equation from the freeze out to the current time.
Although, within $R$-parity
conservation, the neutralino cannot decay, it can pair annihilate or
coannihilate with other supersymmetric particles
into Standard Model particles.
Coannihilation effect is exponentially suppressed by Boltzmann factor, so it is
only important when the mass difference between the particles involved is
small~\cite{coan}.

To understand how the neutralino-stop degeneracy can occur, let us look at the
mass spectrum. 
The lightest neutralino in this model is mostly Bino with mass 
$\simeq 0.4 \, m_{1/2}$. 
The stop square-mass matrix is 

\begin{equation}
\left( \begin{array}{cc}
m_{\tilde{t}_L}^2 &  -m_t(A_{t} +\mu \cot \beta) \\ & \\
 -m_t(A_{t} +\mu \cot \beta) & m_{\tilde{t}_R}^2 \end{array} \right)
\end{equation}

Although $\tilde{t}_R$ is generally heavier than $\tilde{\tau}_R$, the
off-diagonal terms can be very large due to the large top mass $m_t$, and this
suppresses
one of the eigenvalues, $m^2_{\tilde{t}_1}$. For a large $|A_t|$ with ${\rm
sign}(A_t) = {\rm sign}(\mu)$, we get a light $\tilde{t}_1$.  

Processes involved in the neutralino-stop coannihilation are listed
in Table 1. The most dominant contribution comes from $\tilde{t}_1
\tilde{t}_1^{\ast} \rightarrow gg$. The $\tilde{t}_1 \tilde{t}_1^* \rightarrow
t \bar{t}$, $\tilde{t}_1 \tilde{t}_1 \rightarrow t t$ and $\chi \tilde{t}_1
\rightarrow t g$ are also dominant when the final states are kinematically
available. Processes that have only electroweak couplings, i.e. $\tilde{t}_1
\tilde{t}_1^* \rightarrow l \bar{l}, W^+ W^-, Z \gamma, ZZ$ and $\chi
\tilde{t}_1 \rightarrow t \gamma$, contribute only $\sim 1 \%
$ and therefore can be neglected.

To see the significant of this coannihilation, we plot the separate contribution
to the annihilation cross section $\sigma_{\rm eff}$ versus $\Delta M \equiv
(m_{\tilde{t}_1} -
m_{\chi})/m_{\chi}$. Fig.~1 shows such a plot with $\tan \beta = 10$, $A_0 =
1000$~GeV and $m_0 = 300$~GeV. For small $\Delta M$, $\sigma_{\rm eff}$ is
dominated by stop-stop ($\tilde{t}_1 - \tilde{t}_1$, $\tilde{t}_1 -
\tilde{t}_1^{\ast}$ and $\tilde{t}_1^{\ast} -
\tilde{t}_1^{\ast}$) annihilation. Notice that, had we neglected the
coannihilation effect, we would get a $\sigma_{\rm eff}$ up to four order of
magnitude less. Passing over the $t g$ production 
threshold at
$\Delta M \simeq 0.2$, the  $\tilde{t}_1 - \chi$ coannihilation becomes dominant.
The coannihilation effect is suppressed when $\Delta M$ is large and, in this
case, can be neglected for $\Delta M \stackrel{>}{\sim} 0.4$.

\section{THE CMSSM PARAMETER SPACE}
For $A_0 = 0$ the stop mass is much heavier than the neutralino mass.
We start seeing the neutralino-stop coannihilation tail at $A_0
\simeq 500$~GeV
for $\tan \beta = 10$. Here we show an example of CMSSM region with
neutralino-stop coannihilation tail.

Fig.~2 is a contour plot on the $(m_{1/2}, m_0)$ plane with
 $\tan \beta = 10$, $A_0=2000$~GeV and $\mu > 0$. The red shaded region is
excluded because it has charged LSP, either stau for small $m_0$ or stop for
small $m_{1/2}$. The light (turquoise) shaded region is the one that has $0.1
\leq \Omega h^2 \leq 0.3$.  
We still see the neutralino-stau coannihilation tail extending to larger
$m_{1/2}$.
The bulk region, common in the CMSSM, is hidden inside the red shading on the
left-bottom corner.
Extending to higher $m_0$ is the neutralino-stop coannihilation tail.

\begin{figure}[htb]
\vspace{9pt}
\includegraphics[height=17pc]{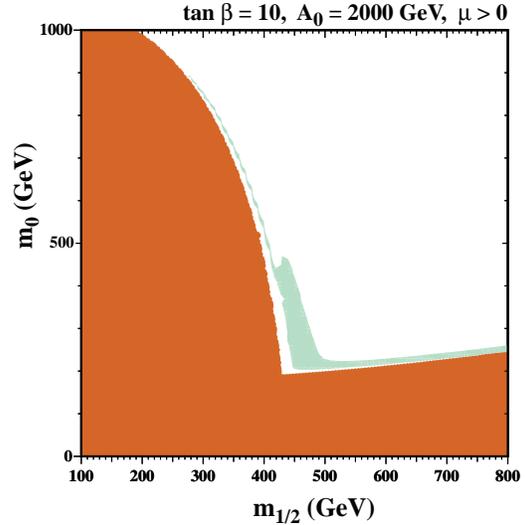}
\caption{The $(m_{1/2}, m_0)$ plane for $\tan \beta = 10$, $\mu > 0$ and $A_0 = 
 2000$~GeV.
The very dark (red) shaded regions are excluded because the LSP is 
the ${\tilde t_1}$ or the ${\tilde \tau_1}$. 
The light (turquoise) shaded regions are those favored by cosmology,
with \protect\mbox{$0.1\leq \Omega h^2 \leq 0.3$} after the inclusion of 
coannihilation effects. 
}
\label{fig:10-2}
\end{figure}

Around $m_{1/2} \simeq 440$~GeV there is top production threshold, and
there the region favored by cosmology is broader.
This can be understood as the following. Beyond the threshold, neutralino can
pair annihilates into $t \bar t$ through channels shown in Fig.~3. Because of
the
large $m_t$, a small \mbox{Higgsino} content in the neutralino is enough to
make this
channels significant. Furthermore, near the neutralino-stop degeneracy line,
$\tilde{t}_1$ is relatively light, and this enhances the cross section to the
level needed to bring the relic density into the preferred range. 

\begin{figure}[htb]
\begin{center}
\setlength{\unitlength}{0.01in}
\begin{picture}(100,130)(0,0)
\put(20,120){\line(1,-1){30}}
\multiput(50,90)(0,-5){10}{\line(0,-1){3}}
\put(20,10){\line(1,1){30}}
\put(50,40){\line(1,-1){30}}
\put(50,90){\line(1,1){30}}
\put(1,120){ $\chi$}
\put(1,5){  $\chi$}
\put(85,120){  $t$}
\put(85,5){  $\bar t$}
\put(55,60){  $\tilde{t}_{1,2}$}
\end{picture}
\end{center} 
\caption{Neutralino pair annihilation diagram to $t \bar{t}$ through  t- and
u-channel stop exchange.}
\end{figure}
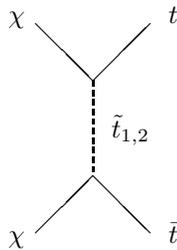

In doing this analysis we also consider experimental constraints, such as the
LEP Higgs, 
 chargino 
 and selectron mass limits, the $b
\rightarrow s \gamma$ branching ratio 
 and the muon anomalous magnetic moment
.  
(Please see Ref.~\cite{eos} for more details on these constraints.)
At this moment the
cosmological region
shown in Fig.~2 is not yet excluded by any of these constraints.
Thus a complete proposal for the future accelerator search for
supersymmetry should also consider the neutralino-stop coannihilation region.

Higher $m_0$ supresses the neutralino-proton cross section. Thus the
neutralino-stop coannihilation predicts the possibility of a lower neutralino
direct detection rate.
However this reduction should be less than about one order of magnitude, as the
tail extending not to a very high $m_0$.  

\section{CONCLUSION}
We have presented the neutralino-stop coannihilation in the CMSSM. This
coannihilation is important when $|A_0| \neq 0$ and large. 
It opens up higher $m_0$ in the parameter space. 
There are now five generic cosmological regions recognized in this model, i.e
the bulk region, the neutralino-stau coannihilation tail, the $H$ and $A$ rapid
annihilation funnel, the focus point channel and the neutralino-stop
coannihilation tail.
Future experiments will exclude more parameter space and we will see which
region will survive. We hope that, ultimately supersymmetry will be
discovered and its parameters will be measured. Before that happens, however, we
should include all possibilities in our consideration.


\begin{thebibliography}{9}
\bibitem{others}
There are many other old and recent papers on the neutralino dark matter in the
CMSSM that are not listed here due to the limited space. For a
more complete list of references please look e.g. Ref.~\cite{eos,eos2}.
\end{thebibliography}
\end{document}